\documentclass[runningheads]{svmult}

\usepackage{graphicx}  
\usepackage{physprbb}  

\begin{document}

\title*{Formation of Globular Clusters: \protect\newline
        In and Out of Dwarf Galaxies}

\toctitle{Formation of Globular Clusters: \protect\newline
          In and Out of Dwarf Galaxies}

\titlerunning{Formation of Globular Clusters}

\author{Oleg Y. Gnedin}

\authorrunning{Oleg Y. Gnedin}

\institute{Space Telescope Science Institute, 3700 San Martin Drive,
           Baltimore, MD 21218, USA; ognedin@stsci.edu}

\maketitle

\begin{abstract}
Despite the rapid observational progress in the study of young massive
star clusters, the formation of globular clusters still remains poorly
understood.  Yet, it is emerging that globular cluster formation is
intimately linked to the formation of the Galaxy.  I discuss a generic
scenario of the cluster formation within progenitor galaxies, based on
the available observational constraints.  The oldest clusters formed
around redshift z = 7, but the process continued at least until z = 3.
Because of their high density, globular clusters survived when their
progenitor galaxies were disrupted by the Galactic tidal field.
\end{abstract}

\section{Observational Constraints}

Building a successful physical model of globular cluster formation has
proved to be an elusive topic, as revealed by the multitude of effort
and the lack of result in the last 30 years.  In essence, this is the
point where the large-scale galaxy formation problem meets the
small-scale star formation problem.  In order to try to overcome the
uncertainties associated with both, we should seek as many clues as
possible from observations.

Fortunately, recent observational studies of star-forming regions in
the Galaxy and young star clusters in interacting galaxies have
improved our understanding of star formation.  The majority of stars
forms in clusters and associations of various sizes.  The hierarchy
ranges from the young OB associations to the old massive globular
clusters, with no special scale between 10 and $10^6\, M_\odot$:
$dN/dM \propto M^{-\alpha}$, $\alpha = 1.5 - 2$ \cite{EE:97,WBM:00}.
Thus globular clusters seem to represent the most massive end of the
general hierarchy.

The observed luminosity function (LF) differs from the {\it initial}
one due to the dynamical evolution.  Small-mass clusters ($M < 10^5\,
M_\odot$) are gradually destroyed by stellar two-body relaxation and
tidal interactions \cite{S:87,GO:97,MW:97,GLO:99}, while very massive
clusters sink to the center via dynamical friction.  However, most of
the high-mass clusters are essentially unaffected by the evolution,
and therefore, preserve the shape of the initial LF.

Observations and modeling of old Galactic and extra-galactic globular
clusters reveal similar general properties that depend little on the
vastly different environment of the host galaxies (small and large,
spiral and elliptical):

\begin{itemize}
\item
{\bf Age.} The oldest globular clusters are old, but realistically
old: $t_{GC} \approx 12.5$ Gyr, with the 95\% probability range $10.2
- 15.9$ Gyr \cite{KC:02}.  There is no problem now with the stars
being older than the Universe, and their age can be used to place
constraints on the epoch of globular clusters formation:
$$
t_f = t_H - t_{GC},
$$
where $t_H \approx 13.5$ Gyr is the Hubble time for the concordance
cosmological model \cite{BOPS:99}.  With all the associated
uncertainties, oldest clusters must have formed between redshifts $z_f
= 7$ and 3 \cite{GLR:01}.

\item
{\bf Metallicity.}  A large spread of metallicity among old clusters,
from 0.01 to 0.3 $Z_\odot$, with a median $\sim 0.03 \, Z_\odot$.
The material from which globular clusters formed must have been
pre-enriched by heavy elements.  A narrow range of metallicity within
individual clusters ($\delta\mbox{[Fe/H]} < 0.1$) indicates that
each globular cluster formed in a single burst.

\item
{\bf Distribution.} Spherically symmetric distribution and little
rotation of the Galactic globular cluster system point to a different
origin than most stars in the rotationally-supported disk.  Also, the
efficiency of globular cluster formation is universally low
\cite{M:99}: the ratio of the mass of globular clusters to the total
baryonic (stellar + gaseous) mass of their host galaxy is the same in
giant and dwarf ellipticals, $\epsilon_{GC} \equiv M_{GC}/M_{\rm bar}
\approx 0.0026 \pm 0.0005$.

\end{itemize}

\section{A generic hierarchical scenario}

Early models of globular cluster formation \cite{PD:68,FR:85,HP:94}
focused on local physical processes, assuming a static environment of
the present Galaxy.  In the last decade the paradigm of galaxy
formation has changed significantly.  We now have good evidence that
galactic halos form as a result of gravitational growth and
interaction of primordial fluctuations within the cold dark matter
model \cite{O:93}.  Small objects collapse first and merge into larger
systems, extending the hierarchy to progressively higher masses.  The
present halo of the Milky Way formed in dozens of mergers of smaller
progenitors \cite{Moore:99,Bullock:01}.

At redshift $z_f \sim 7$, globular clusters {\it must} form within
small progenitor galaxies, because large galaxies do not exist yet.
The primordial gas would cool by the hydrogen recombination line
emission within the first halos with the virial temperature $T_{\rm
vir} > 10^4$ K and settle into a rotationally-supported,
self-gravitating disk.  Local instabilities would break the disk into
dense clumps, which could further fragment into stars.  Star formation
in the local universe proceeds in thin disks, and it is natural to
assume the same for the high-redshift progenitors.  The overall
spherical geometry of the globular cluster system follows from the
subsequent mergers and tidal disruption of the progenitor galaxies.

For illustration, consider a simple scenario of the first episode of
star formation in a progenitor galaxy.  Assume that the scale-free
mass function of giant molecular clouds translates into a similar
cluster mass function, $dN_{cl}/dM \propto M^{-\alpha}$.  For $\alpha
< 2$, the total mass of the star cluster hierarchy within a progenitor
is dominated by the most massive cluster, $M_{\rm max}$.  Combined
with the observed formation efficiency $\epsilon_{GC}$, the total
baryon mass of the progenitor could be estimated as $M_{\rm
max}/\epsilon_{GC}$.  For a cosmic baryon fraction, $f_b =
\Omega_B/\Omega_0 \approx 0.13$, this corresponds to the virial mass
of the progenitor $M_{\rm vir} = M_{\rm max}/(f_b \, \epsilon_{GC})
\approx 3000 \, M_{\rm max}$.

\begin{figure}[t]
\begin{center}
\includegraphics[width=.7\textwidth]{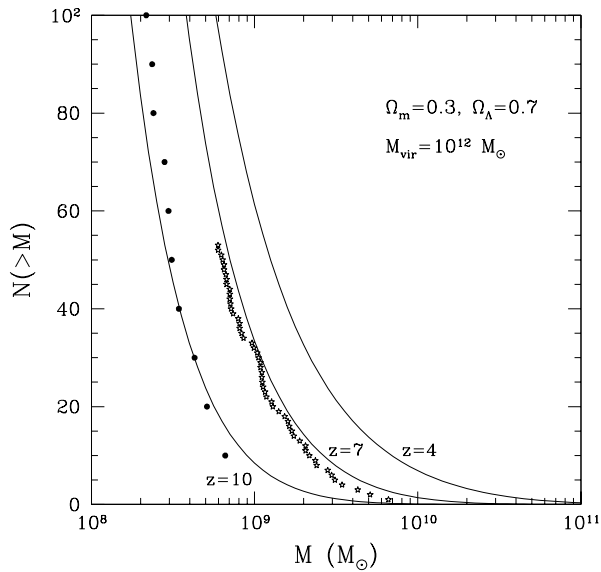}
\end{center}
\caption[]{Cumulative number of progenitors of the present Milky Way
halo more massive than $M$ at three different redshifts.  Overplotted
by asterisks is the luminosity function of the massive Galactic globular
clusters rescaled by a factor $(f_b \, \epsilon_{GC})^{-1} \approx
3000$, for a constant mass-to-light ratio $M/L_V = 2$.
If some progenitors had higher formation efficiency $\epsilon_{GC}$
and others did not form globular clusters at all, the distribution
would shift towards higher redshift.  For example, filled circles
show the mass function in the case when 10\% of the progenitor halos
form clusters with the efficiency $10 \, \epsilon_{GC}$.}
\label{fig:1}
\end{figure}

Figure \ref{fig:1} shows the distribution of the progenitor halos of
the Milky Way calculated using the extended Press-Schechter formalism
\cite{LC:93}.  At redshift $z = 7$, the distribution fits accurately
the high end of the LF of the Galactic clusters renormalized by $(f_b
\, \epsilon_{GC})^{-1}$.  In this model, the most massive cluster
originates in the progenitor with the mass $M_{\rm vir} \approx
6\times 10^9\, M_\odot$ and circular velocity $v_c \approx 25$ km
s$^{-1}$.  Correspondingly, smaller progenitors host smaller globular
clusters.

\begin{figure}[t]
\begin{center}
\includegraphics[width=0.48\textwidth]{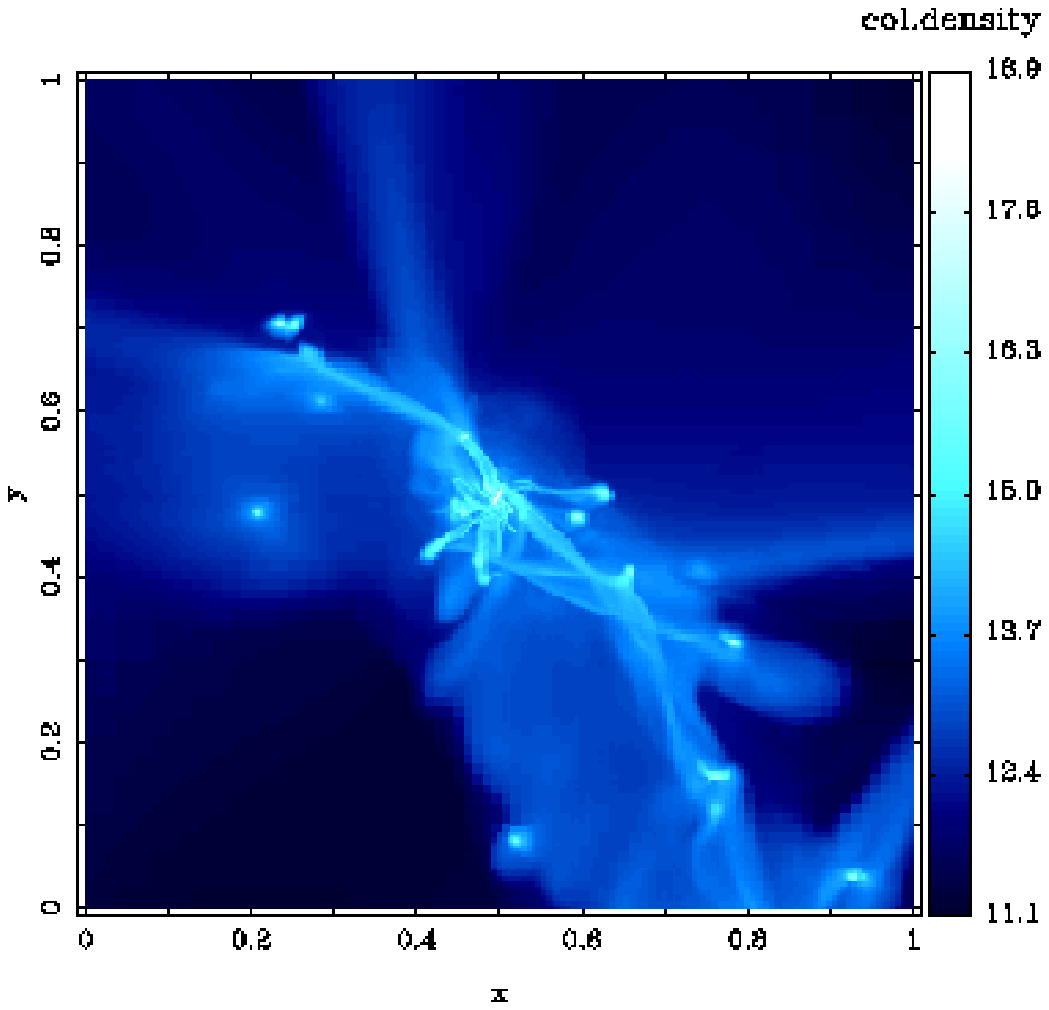}\hspace{0.3cm}
\includegraphics[width=0.48\textwidth]{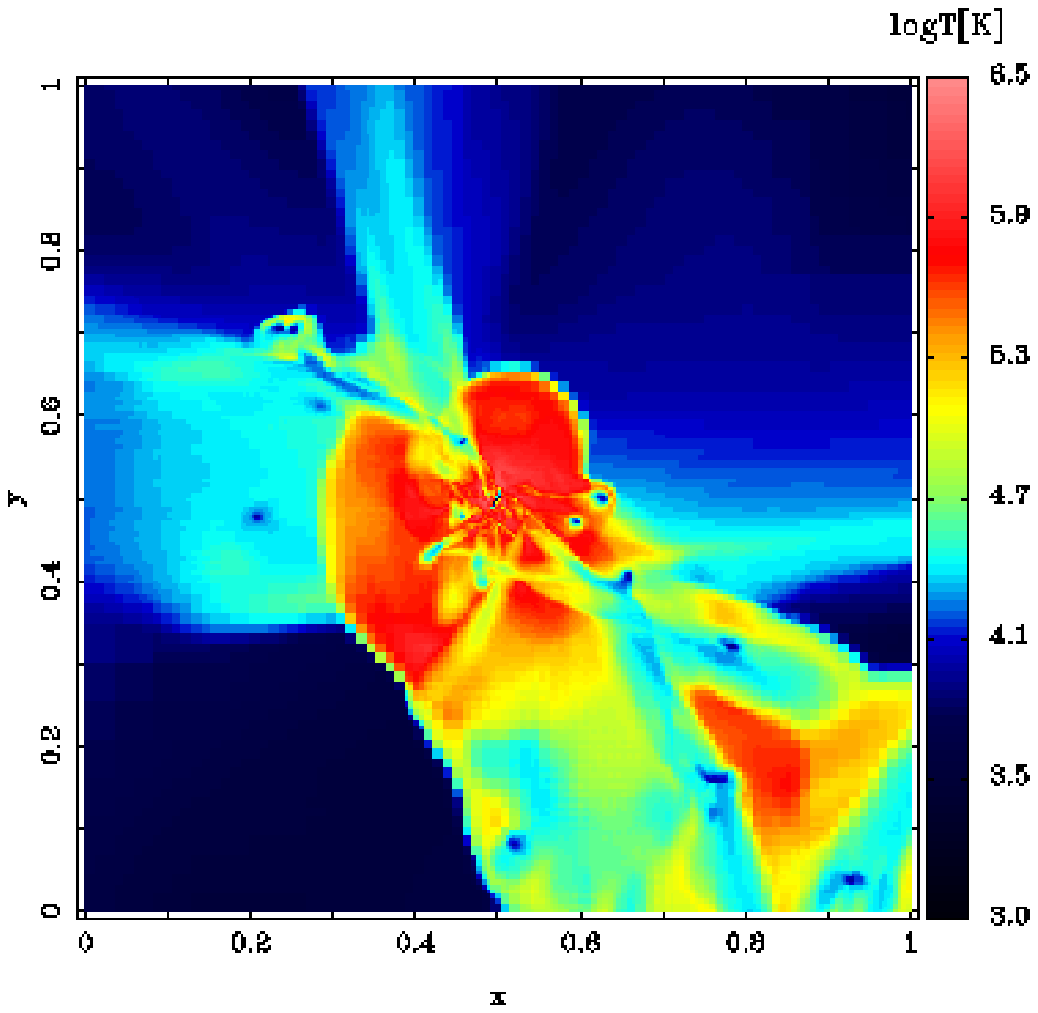}
\vspace*{0.2cm}
\includegraphics[width=0.48\textwidth]{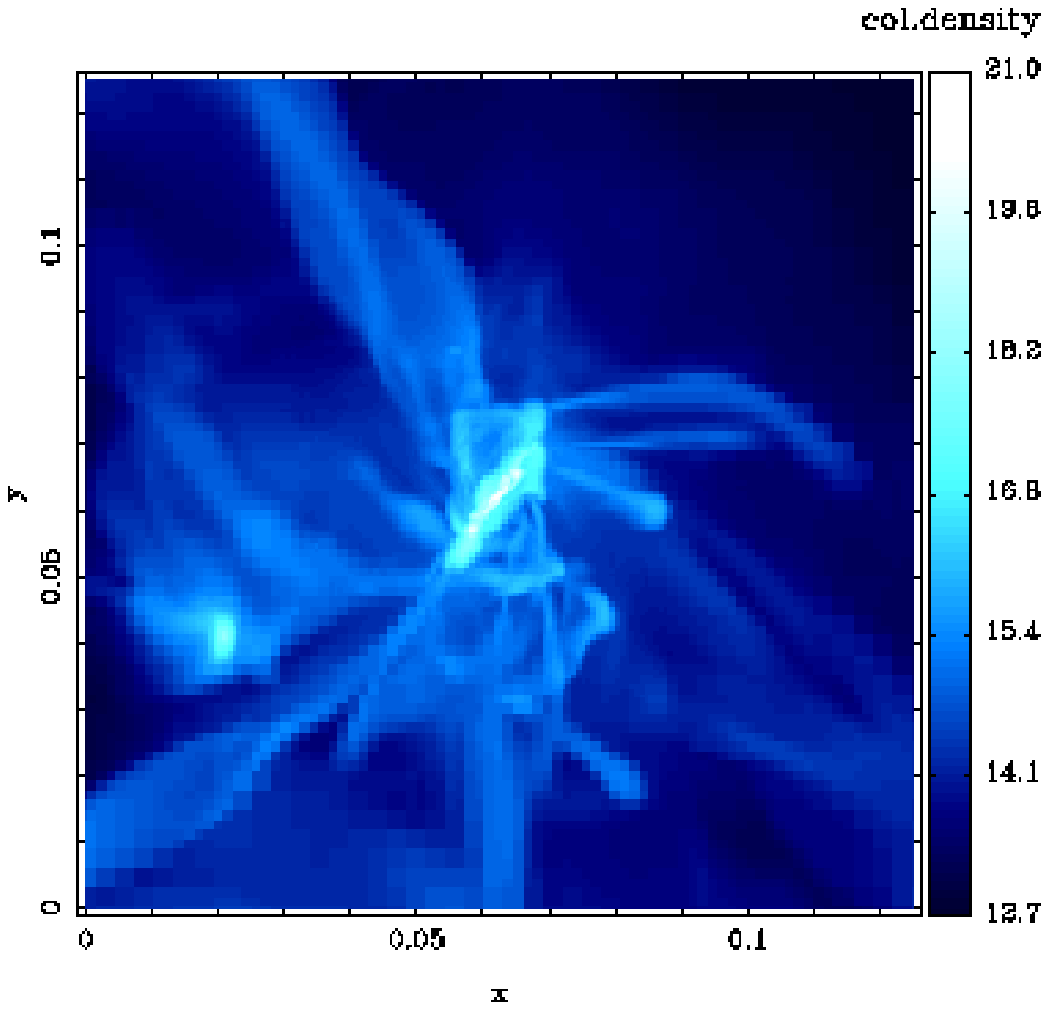}\hspace{0.3cm}
\includegraphics[width=0.48\textwidth]{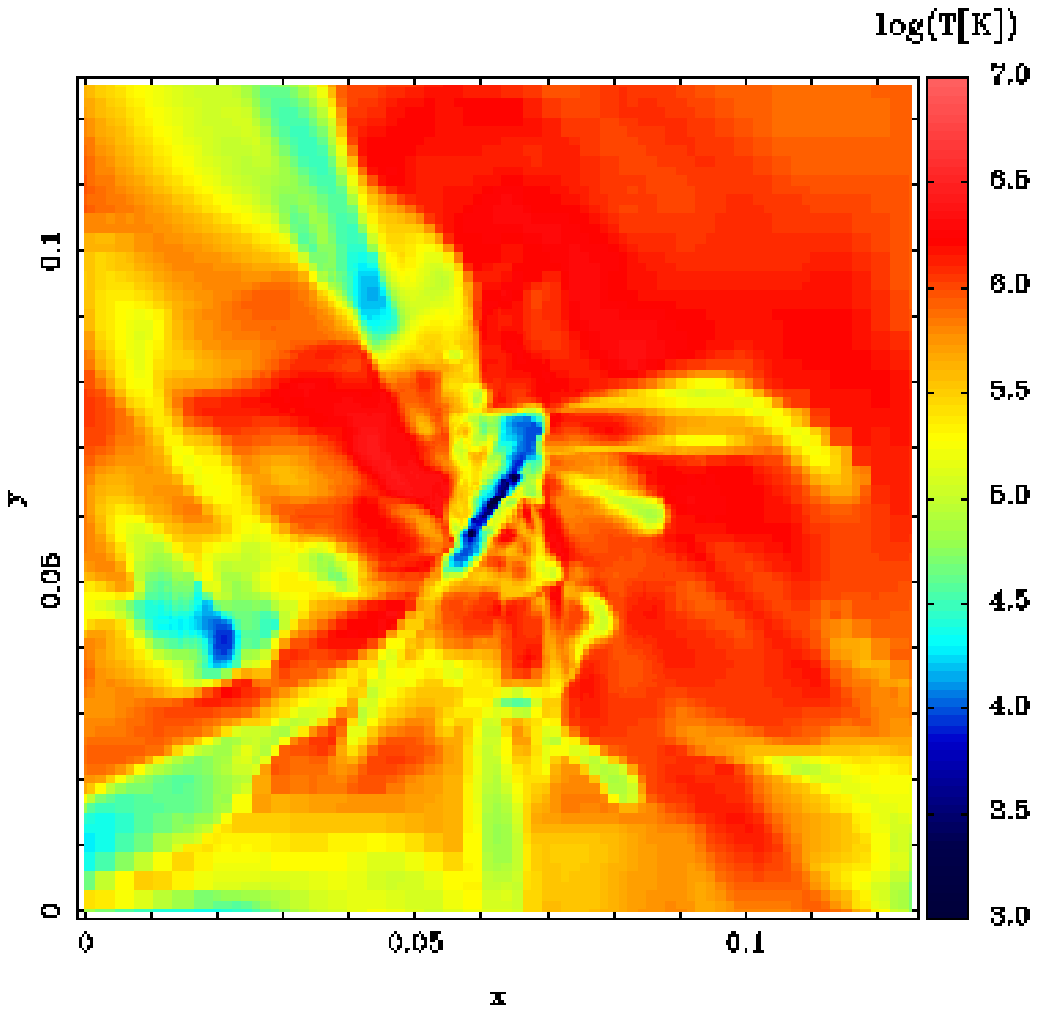}
\end{center}
\caption[]{ART simulation of the disk galaxy formation at $z=4$ (by
A. Kravtsov).  Upper panels show the gas column density and
the temperature in a box of 1 $h^{-1}$ Mpc comoving.  The lower panels
show the central 125 $h^{-1}$ kpc around the cold disk.}
\label{fig:2}
\end{figure}

\begin{figure}[t]
\begin{center}
\includegraphics[width=.8\textwidth]{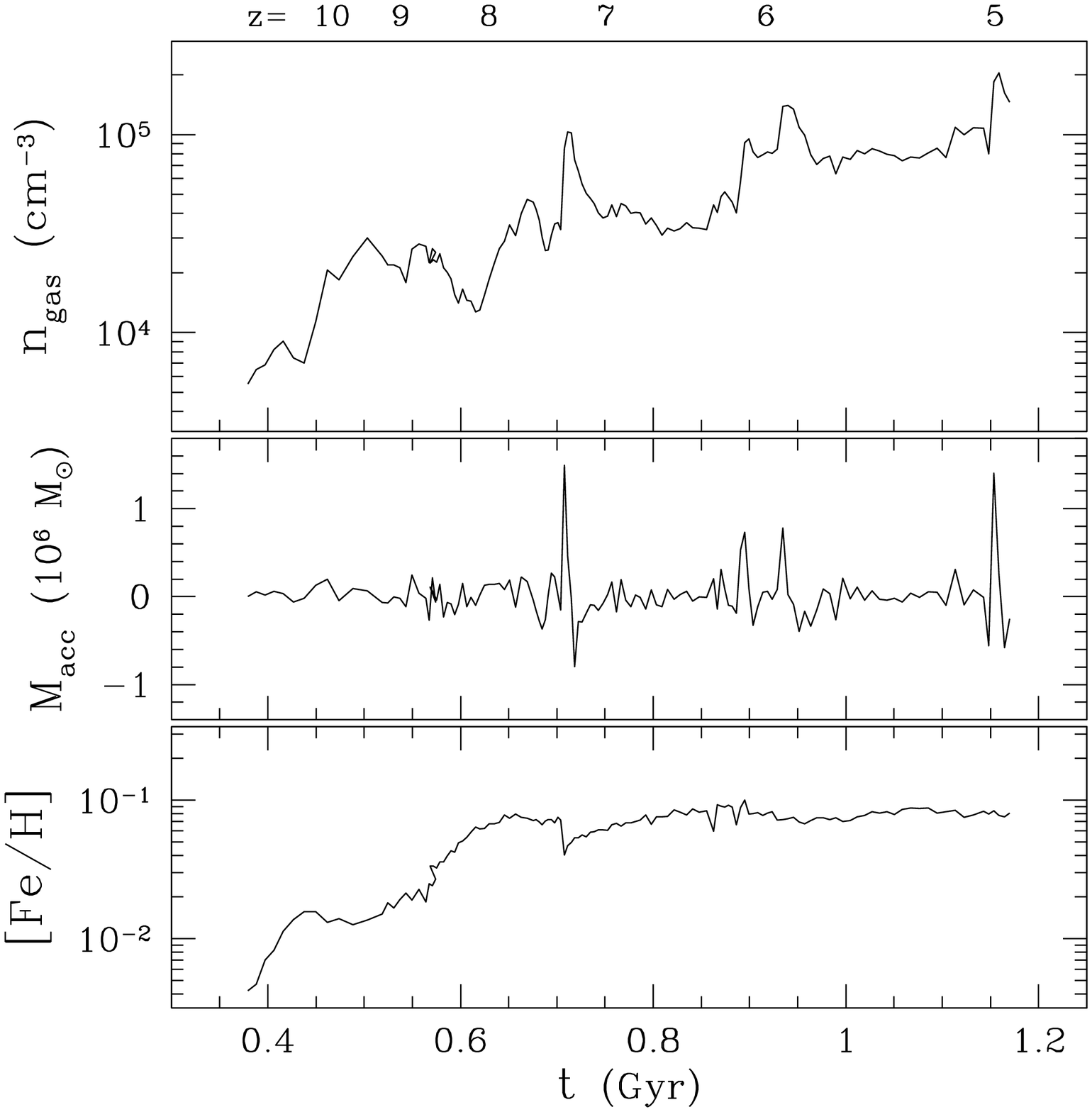}
\end{center}
\caption[]{ART simulation: the density, the mass accreted between
successive time steps, and the metallicity of gas in the central 20 pc
of the main progenitor.}
\label{fig:3}
\end{figure}

Dwarf galaxies in the Local Group provide evidence that globular
clusters can form in such low potential wells.  Even the Fornax dwarf
spheroidal galaxy with the mass of only $2\times 10^7\, M_\odot$
contains 5 clusters, while the LMC is forming star clusters now.
Perhaps the most striking example is $\omega$ Cen, the most massive
Galactic globular cluster.  Unlike all other globular clusters, it
displays multiple stellar populations with a significant spread of age
($3-5$ Gyr) and metallicity ($\delta[Fe/H] \sim 1$).  Self-enrichment
is unlikely, as this cluster is not special among others in its age
and the ability to retain its own stellar winds \cite{Getal:02}.  If
the oldest stars in $\omega$ Cen had formed at its present position
close to the center of the Galaxy (the pericenter is 1.2 kpc), the
cluster would be stripped of any accumulated gas every $6 \times 10^7$
yr and would be unable to form a second generation of stars.  The only
viable scenario is that it formed within a progenitor galaxy at a
large distance and was brought towards the center on a very eccentric
orbit.

\section{Resolving globular clusters in cosmological simulations}

The complexity of gas dynamics in the high-redshift progenitors can
only be probed with direct numerical simulations.  They must involve a
large range of scales, from the whole galaxy to the individual
clusters, and include proper tidal interactions among the progenitors.
To this end, in collaboration with A. Kravtsov, we are analyzing the
results of a very high-resolution simulation of the formation of a
Milky Way-type galaxy.  The simulation is done using the Adaptive
Refinement Tree (ART) code \cite{ART}, which can refine the resolution
grid in the high-density regions by a factor $2^9$.

Figure \ref{fig:2} shows the distribution of gas centered on the most
massive progenitor at $z=4$.  The gas flows inward along the filaments
and is shock-heated when it enters the dense virialized halo.  A big
red corona around the center in the upper right panel is the hot gas
at the virial temperature.  The lower panels show a thin disk which is
actively forming stars in the middle; at $z=4$ its physical size is
$2-4$ kpc.

In the central part of the disk the gas is not allowed to cool below
300 K, otherwise the density becomes too high and the simulation
halts.  This is a numerical limitation of the run which can be
overcome in the future.  Even with that criterion, the density in the
central resolution cell exceeds $10^5$ cm$^{-3}$.  This is close to
the density of molecular clouds and such an object is very likely to
form a dense star cluster.  How do we determine when a globular
cluster forms?

Figure \ref{fig:3} shows the growth of density in the central cell,
which has several noticeable peaks associated with the fast episodes
of accretion.  Note on the lower right panel of Fig. \ref{fig:2} that
the filaments stretching all the way to the center bring in some gas
that is so dense that even if shocked, it quickly cools back.  Thus
the central region is directly accreting cold, metal-poor gas.  Such
episodes of accretion or merging are apparent in the middle panel of
Fig. \ref{fig:3}, which shows the mass added between successive time
steps.  Some of the gas is ejected from the center by the
hydrodynamical flows, so the mass difference can be negative.

The two episodes at $z \approx 7.3$ and $z \approx 5.9$ are most
likely to form massive star clusters.  The accreted gas suddenly
increases the external pressure on the molecular clouds and triggers
their collapse.  The time for cluster formation is short, after
$30-50$ Myr the remaining gas may be thrown out by a new merger.
After redshift of 5 the conditions become less favorable -- the
accretion slows down and the accumulated stellar mass starts to affect
gas dynamics.

The bottom panel of Fig. \ref{fig:3} shows the evolution of
metallicity due to the SN type II ejecta.  It quickly rises to 10\% of
solar and then varies slowly, either up or down.  Note that the sudden
events of accretion of the cold, unprocessed gas lowers the mean
metallicity even in the very central region.  If a series of globular
clusters forms between $z=8$ and 5, the younger ones are not
necessarily more metal-rich than the older ones.  Thus we should not
expect a clear age-metallicity correlation for the clusters formed at
high redshift.

We will continue the investigation of the gas dynamics and star
formation in all progenitor halos, with the goal of reproducing the
present ages, metallicity, and spatial distribution of the globular
clusters.

\end{document}